\documentclass[proof]{WileyASNA-v1}

\articletype{}%

\received{}
\revised{}
\accepted{}

\begin{document}

\title{Equatorially trapped Rossby waves in radiative stars}

\author[1,2,3]{M. Albekioni*}

\author[4,2,3]{T.V. Zaqarashvili}

\author[2,3]{V. Kukhianidze}

\author[1,2,3]{E. Gurgenashvili}

\author[4]{P. Bourdin}

\authormark{AUTHOR ONE \textsc{et al}}

\address[1]{\orgdiv{Institut f\"ur Astrophysik}, \orgname{Georg-August-Universit\"at}, \orgaddress{\state{Friedrich-Hund-Platz 1, 37077, G\"ottingen}, \country{Germany}}}

\address[2]{\orgdiv{Department of Astronomy and Astrophysics at Space Research Center, School of Natural Sciences and Medicine}, \orgname{Ilia State University}, \orgaddress{\state{Kakutsa Cholokashvili Ave. 3/5, Tbilisi 0162}, \country{Georgia}}}

\address[3]{\orgdiv{Evgeni Kharadze Georgian National Observatory}, \orgaddress{\state{Abastumani, Adigeni 0301}, \country{Georgia}}}

\address[4]{\orgdiv{Institut of Physics, IGAM}, \orgname{University of Graz}, \orgaddress{\state{Universit\"atsplatz 5, 8010 Graz}, \country{Austria}}}

\corres{*Mariam Albekioni \email{mariam.albekioni.1@iliauni.edu.ge}}

\presentaddress{}

\abstract{Observations by recent space missions reported the detection of Rossby waves (r-modes) in light curves of many stars (mostly A, B, and F spectral types) with outer radiative envelope. This paper aims to study the theoretical dynamics of Rossby-type waves in such stars. Hydrodynamic equations in a rotating frame were split into horizontal and vertical parts connected by a separation constant (or an equivalent depth). Vertical equations were solved analytically for a linear temperature profile and the equivalent depth was derived through free surface boundary condition. It is found that the vertical modes are concentrated in the near-surface layer with a thickness of several tens of surface density scale height. Then with the equivalent width, horizontal structure equations were solved, and the corresponding dispersion relation for Rossby, Rossby-gravity, and inertia-gravity waves was obtained. The solutions were found to be confined around the equator leading to the equatorially trapped waves. It was shown that the wave frequency depends on the vertical temperature gradient as well as on stellar rotation. Therefore, observations of wave frequency in light curves of stars with known parameters (radius, surface gravity, rotation period) could be used to estimate the temperature gradient in stellar outer layers. Consequently, the Rossby mode may be considered as an additional tool in asteroseismology apart from acoustic and gravity modes.}


\maketitle

\section{Introduction}
Rossby (planetary) waves govern large-scale dynamics of rotating spheres\footnote{The waves are mostly known as r-modes in stellar physics \citep{Papaloizou1978}}. The waves have been studied for centuries starting from \citet{Hadley1735}. They are characterized as low-frequency waves compared to the rotation frequency of the sphere and the solutions were found by Hough \citep{Hough1897, Hough1898} based on Laplace tidal equations \citep{Laplace1893}. The existence of Rossby waves is associated with the Coriolis force and conservation of absolute vorticity on a rotating sphere. The first observational description and physical basics of the waves were discussed by \citet{Rossby1939} in the Earth context during his work on a global weather forecast at the Massachusetts Institute of Technology. Consequently, the Rossby waves are well studied in Earth atmosphere and oceans by observations and theory  \citep{Hovmoller1949, Eliasen1965, Yanai1983, Lindzen1984, Hirooka1989, Madden2007, Chelton1996, Hill2000, Haurwitz1940, Lindzen1967, Gill1982, Pedlosky1987, Platzman1968, Salby1984}. 

Rossby waves have an important role in the dynamic of different astrophysical objects like accretion disks, solar system planets, stars, etc. \citep{Zaqarashvili2021}. The waves have been recently detected on the Sun by granular tracking and helioseismology  \citep{Lopten2018, Liang2019, Hanasoge2019, Proxauf2020, Gizon2021, Hanson2022} as well as by coronal bright points \citep{McIntosh2017, Krista2017}. 
It is also suggested that the Rossby waves may lead to intermediate periodicity in solar activity \citep{Zaqarashvili2010, Gurgenashvili2016, Gurgenashvili2017, Dikpati2020}. Theoretical description of Rossby waves in the solar interior and tachocline is also significantly developed in recent years \citep{Zaqarashvili2007, Zaqarashvili2018, Gachechiladze2019, Dikpati2018, Dikpati2022, Gizon2020, Bekki2022, Horstmann2023}. 

Huge observational data collected by recent space missions TESS (Transiting Exoplanet Survey Satellite), CoRoT (Convection, Rotation, and planetary Transits), and Kepler led to the detection of Rossby waves in light curves of stars. First, \citet{VanReeth2016} reported the evidence of r-modes in the Gamma Doradus stars using Kepler data. Then the waves have been observed in many stars with different spectral classes \citep{Saio2018, Li2019, Jeffery2020, Samadi2020, Takata2020, Saio2022}. It has been also suggested that the Rossby waves may lead to the observed short-term stellar cycles \citep{Lanza2009, Bonomo2012, Gurgenashvili2022}.

The theory of Rossby waves was developed in the context of the Earth's atmosphere and oceans mostly in the shallow water approximation, which considers the shallow layer of the sphere with a homogeneous density. This approximation is valid for the layers, which have smaller widths compared to the density scale height. On the other hand, stellar interiors generally cover many scale heights especially upper parts with lower temperatures. Therefore, the consideration of the radial direction is necessary to model the Rossby waves in stars. For this reason, the separation of horizontal and vertical equations allow us to find the solutions of Rossby waves in some approximations. Using expansions with small parameters, \citet{Papaloizou1978} studied the high order r-modes in 3D spherical geometry of rapidly rotating stars, while \citet{Provost1981} and \citet{Saio1982} analyzed the low order modes in the slow rotation limit. However, it was shown by \citet{Taylor1936} that the dynamics of Rossby waves in stratified fluids are identical to the waves in a homogeneous layer that has a width of the equivalent depth corresponding to the separation constant of horizontal and vertical equations. Then, the Rossby waves in stratified fluids (such as stellar interiors) can be described in shallow water approximation if the separation constant is known. \citet{Townsend2003} used the spherical hydrodynamic equations in the traditional approximation and found the relation between wave frequency and separation constant (see also in \citet{Lee1997}). It was found that the solutions of horizontal equations (in fact Laplace tidal equations) are approximated by parabolic cylinder equation (or Schr\"odinger equation) in the low-frequency limit, which has exact solutions in terms of Hermite polynomials. This phenomenon is well studied in the Earth context by \citet{Longuet-Higgins1968} in spherical coordinates and by \citet{Matsuno1966} in Cartesian coordinates. Namely, the low-frequency waves in shallow water approximation (obtained either in fast rotation or in a narrow width of the layer) are confined around the equator and hence are known as equatorially trapped waves. Equatorially trapped waves have the same solutions and dispersion relations in spherical and Cartesian coordinates \citep{Zaqarashvili2021}, which allows us to consider simpler rectangular symmetry. Recently \citet{Albekioni2023} used the traditional approximation to solve the dynamics of Rossby waves in horizontal and vertical directions in rotating early-type stars with outer radiative envelopes. The horizontal behavior of the waves was found to be governed by a parabolic cylinder equation with the solutions of Hermite polynomials. While the vertical behavior of the waves was governed by the Bessel equation for the linear vertical temperature gradient. Consequently, the vertical solutions were found in terms of Bessel functions and hence the free surface boundary conditions allowed us to find the vertical modes with corresponding separation constant. Then the separation constant was used in horizontal equations to find the frequencies of different wave modes with various horizontal wave numbers. However, \citet{Albekioni2023} found the solutions only for the first vertical mode in a particular temperature gradient, while the formalism allows us to consider also overtone vertical modes and different values of temperature gradients. Vertical overtones lead to the construction of period spacing patterns, which are very important to compare the theory to observations \citep{Aerts2021}, therefore additional study related to vertical overtones is certainly required.

In this paper, we expand the results of \citet{Albekioni2023} for various vertical modes of Rossby waves and different rates of vertical temperature gradients for stars with outer radiative envelopes. 

\section{Theoretical model}\label{sec1}

We use the linear adiabatic hydrodynamic equations in a rotating frame of stellar surface:
\begin{equation}
     {\rho_0}\frac{\partial \vec{v}}{\partial t} + 2 {\rho_0}{\vec\Omega}\times  \vec{v} = - \nabla p' + \rho' \vec {g},
     \label{Eq.(1)}
   \end{equation}
   \begin{equation}
     \frac{\partial \rho'}{\partial t} + (\vec{v}\cdot \nabla)\rho_0 + \rho_0 \nabla \cdot \vec{v} =0,
     \label{Eq.(2)}
     \end{equation}
     
     \begin{equation}
     \frac{\partial p'}{\partial t} + (\vec{v} \cdot \nabla) p_0 + \gamma p_0 \nabla \cdot \vec{v} =0,
     \label{Eq.(3)}
     \end{equation}
where $\rho_0$ ($p_0$) is the equilibrium density (pressure), $\rho'$ ($p'$) is the perturbation of density (pressure), $\vec{v}$ is the perturbation of velocity,  $\vec{g}$ is the gravitational acceleration, $\vec \Omega$ is the angular frequency of rotation and $\gamma=c_p/c_v$ is the ratio of specific heats. We note that solid body rotation is adopted throughout the paper. The effects of differential rotation will be studied in the future. Undisturbed
medium is assumed to be in vertical hydrostatic balance, which with the ideal gas law gives the vertical distribution of undisturbed physical parameters (density, pressure, and temperature). Details of the initial setup and mathematical formalism can be found in \citet{Albekioni2023} and we do not repeat it here.  \\

\begin{figure*}
\centering
\includegraphics[width=18cm]{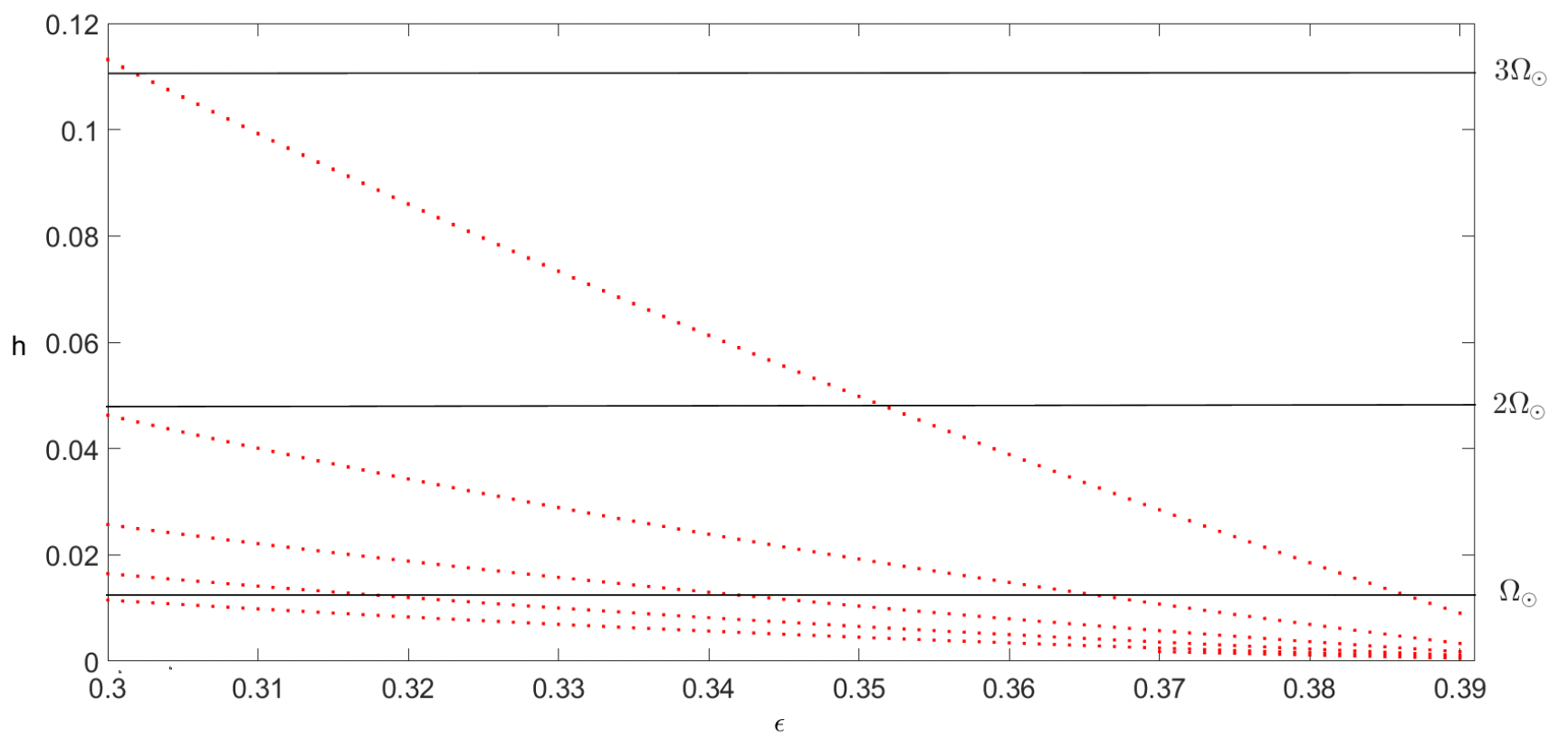}
\caption{Equivalent depth (normalized by the surface scale height, $H_0$) vs the temperature gradient, $\epsilon$, for the first five zeroes of Eq. ~(\ref{Eq.(10)}) i.e. for the first five vertical modes (the first mode corresponds to the upper curve, and the fifth mode corresponds to the lower curve).  Horizontal black lines show the upper limits of $h_c$, which satisfy the polar boundary conditions, for different stellar rotation rates ($\Omega_{\odot}$ is the angular frequency of the Sun). All values of $h<h_c$ for each rotation rate lead to the bounded solutions along the latitudinal direction. 
\label{fig1}}
\end{figure*}

Considering the Cartesian coordinate system, the Fourier transform of the form $e^{i(-\sigma t +k x)}$ and the vertically hydrostatic assumption of perturbed variables, Eqs. (\ref{Eq.(1)})-(\ref{Eq.(3)}) lead to the two equations (see details in \citet{Albekioni2023})
     \begin{equation}
        \frac{\partial^2 \Psi}{\partial y^2} + \left[\frac{\sigma^2 -f^2}{g h} - k^2 - \frac{k}{\sigma}\frac{d f}{dy} \right ] \Psi(y) = 0,
        \label{Eq.(4)}
        \end{equation}
        
        $$ \frac{\partial }{\partial z} \left [  \frac{\gamma H}{1-\gamma(1+H^{\prime})} \frac{\partial }{\partial z} \right ]V(z) - 
$$
\begin{equation}
-\left[ \frac{\gamma(1+H^{\prime})^2}{4H[1-\gamma(1+H^{\prime})]} - \frac{\gamma H^{\prime \prime}}{2[1-\gamma(1+H^{\prime})]^2} + \frac{1}{h} \right]V(z) = 0,
\label{Eq.(5)}
        \end{equation}
where $\Psi(y)$ and $V(z)$ are the functions describing the latitudinal and the radial parts of the latitudinal velocity so that $v_y=\Psi(y)V(z)$, $\sigma$ is the frequency of waves, $k$ is the wavenumber along the $x$ axis, $f=2\Omega \sin \theta$ is the Coriolis parameter ($\theta$ being as latitude) and $H(z)=k_bT(z)/m g$ is the density scale height ($k_b$ is the Boltzmann constant and $m$ is the mass of hydrogen atom). Here $H^{\prime}$ is the first derivative and $H^{\prime \prime}$ is the second derivative with $z$. Note that $x,y,z$ axes are directed towards the rotation, the north pole, and vertically upwards, respectively. $\rho_0$ is absorbed in the expression of velocity so that $v_y$ is multiplied by $\sqrt {\rho_0}$. We used the method of separation of variables to separate the equations into $y$- and $z$-dependent parts, so that $h$ is the separation constant.
The physical meaning of $h$ can be understood for a simple isothermal case, $H=const$.  In this case, Eq. ~(\ref{Eq.(5)}) has periodic solutions when $k^2_z=(\gamma -1)/(\gamma H h)-1/4H^2>0$, which means that $h<4H(\gamma -1)/\gamma$. Then $k_z \sim h^{-1/2}$ i.e. smaller $h$ yields the shorter vertical wavelength. A star with a surface temperature of 10 000 K yields $H=300$ km, then $h$ must be smaller than 480 km in order to have periodic solutions in vertical directions. The corresponding vertical wavelength is a few Mm and less, therefore the waves will have a small vertical extent near the surface but may be increased with depth. 

Eq. ~(\ref{Eq.(4)}) is equivalent to the equation that governs shallow water equatorially trapped waves in a homogeneous layer with the width of $h$ (e.g., \citet{Matsuno1966}). In fact, this is the Taylor theorem \citep{Taylor1936} stating that the dynamics of Rossby waves in stratified fluids are identical to the waves in a homogeneous layer that has a width of equivalent depth, $h$. Therefore, one can use the shallow water equations to describe the Rossby waves in a stratified fluid, if the equivalent depth of the corresponding wave mode is known. Note that the equivalent depth is different for different modes of Rossby waves.

Equations (\ref{Eq.(4)})-(\ref{Eq.(5)}) govern the dynamics of Rossby and inertia-gravity waves in the vertical and horizontal directions. Note that the internal gravity and acoustic waves are neglected from the consideration using low-frequency approximation. This approximation is justified as the acoustic and internal gravity waves have time scales of minutes and hours respectively, while the Rossby and inertia-gravity waves have much longer time scale of the order of days. The equations (\ref{Eq.(4)})-(\ref{Eq.(5)}) can be solved in two different ways. First, one can solve the latitudinal part of the equations and calculate the separation constant based on the boundary conditions at the poles. Then the obtained separation constant can be used to find the radial dependence of solutions through the solution of a system of vertical equations (this approach corresponds to the forced oscillation case in the Earth's atmospheric science). \citet{Lee1997} and \citet{Townsend2003} used the same approach in spherical geometry separating initial equations into two systems of equations with latitudinal and vertical dependence. They solved the latitudinal equations and derived the relations between the wave frequency and the separation constant. Second, one can first solve the radial part of the equations using the vertical boundary conditions and find the corresponding separation constant that is the equivalent depth (this approach corresponds to the so-called free oscillation case in the Earth's atmospheric science). Then one can use the equivalent depth to find the latitudinal dependence and frequency of the waves. \citet{Albekioni2023} used the second approach and found equatorially trapped Rossby waves satisfying the bounded conditions at poles. Here we use the approach as in \citet{Albekioni2023}, hence we will first solve Eq. (\ref{Eq.(5)}) and find the equivalent depth according to the boundary conditions in the vertical direction.

\begin{figure*}
\centering
\includegraphics[width=18cm]{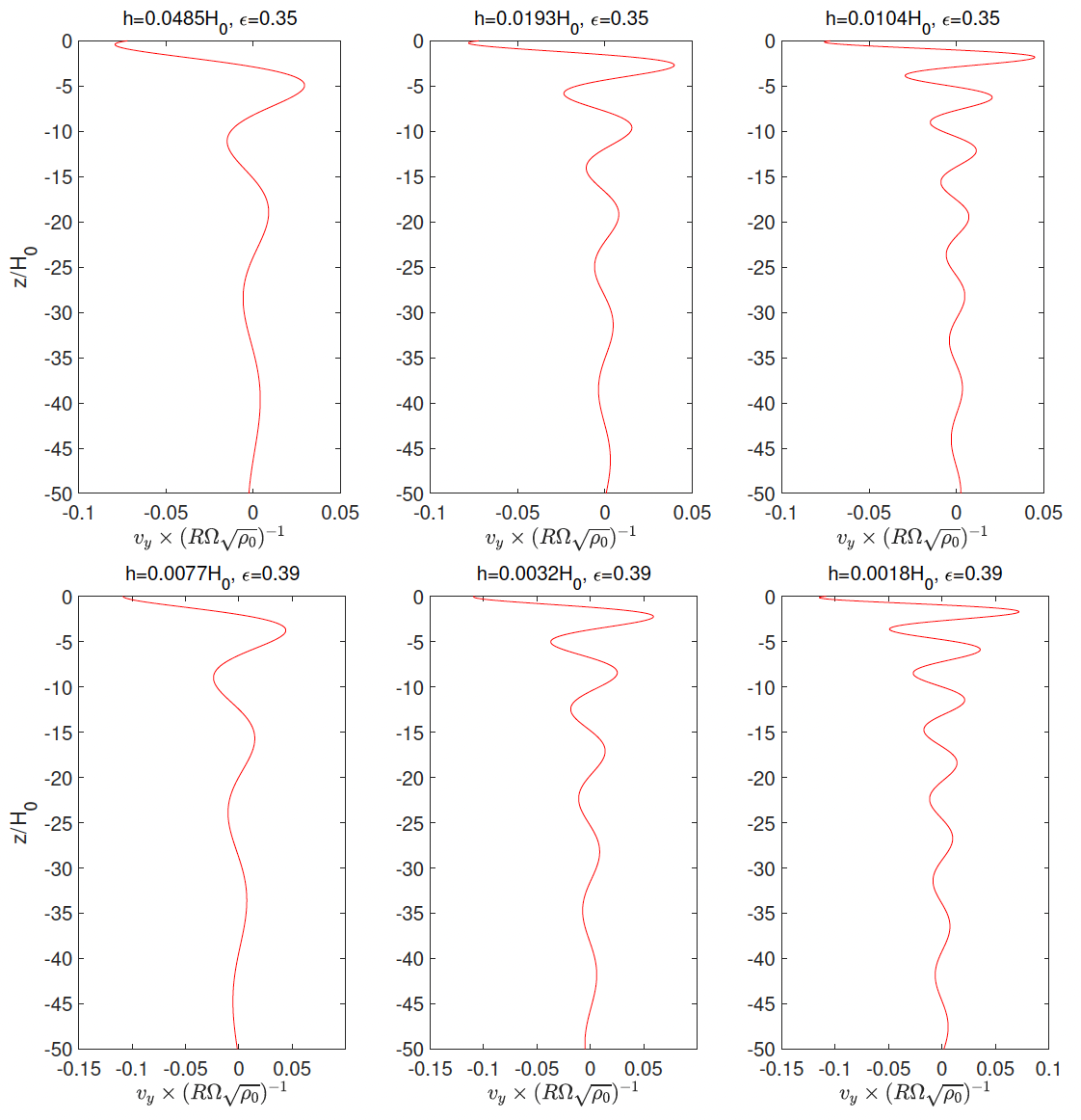}
\caption{Vertical structure of first three vertical modes of Rossby waves for the temperature gradient of $\epsilon=0.35$ (upper panels) and $\epsilon=0.39$ (lower panels). Left, middle, and right panels correspond to the first, second, and third modes of Rossby waves, respectively. Note that $\rho_0$ is absorbed in the expression of velocity in the used formalism so that $v_y$ is multiplied by $\sqrt {\rho_0}$. Therefore, the velocity plotted in this figure is non-dimensional. \label{fig2}}
\end{figure*}

\section{Vertical structure of Rossby waves}\label{sec2}

Solutions of Eq. ~(\ref{Eq.(5)}) are determined by the vertical variation of the density scale height, which actually depends on the vertical temperature gradient. In this paper, we use the linear profile of the scale height with a uniform vertical gradient 
   \begin{equation}
 H=H_0-\epsilon z,
  \label{Eq.6}
     \end{equation}
which is equivalent to the temperature profile of the form
        \begin{equation}
  T=T_0 \left (1- \epsilon \frac{z}{H_0} \right ),
  \label{Eq.(7)}
     \end{equation}
  where $H_0$ and $T_0$ are the scale height and the temperature at the surface, $z=0$, respectively. 

  In this paper, we consider early-type stars with outer radiative envelope, therefore $\epsilon < (\gamma - 1)/\gamma =0.4$ should be satisfied. The linear temperature gradient assures the existence of a radiative environment through the considered extent of the stellar interior. For other profiles, the vertical temperature gradient generally increases with depth, which unavoidably leads to the violation of the radiative condition at some distance from the surface. It was shown by \citet{Albekioni2023} that the Rossby waves are confined near-surface layers of stars with the thickness of 30-50 $H_0$, hence the linear profile is justified. Therefore, here we only consider an uniform linear profile of the temperature.

\begin{figure*}
\centering
\includegraphics[width=18cm]{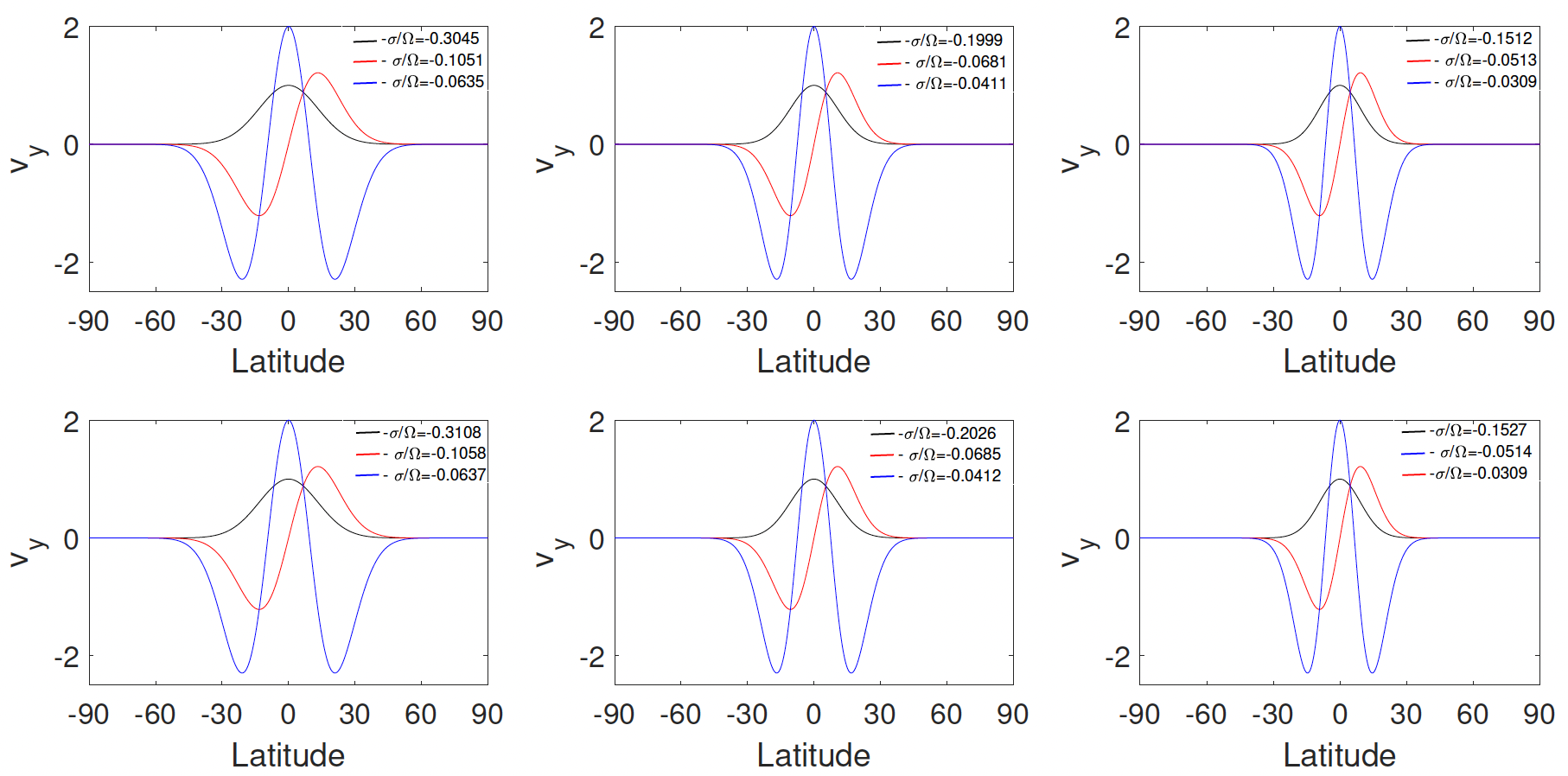}
\caption{Latitudinal structure of Rossby waves for temperature gradient $\epsilon=0.39$, stellar angular frequency of $3\Omega_{\odot}=9 \times 10^{-6} s^{-1}$ (upper panel) and $5\Omega_{\odot}=15 \times 10^{-6} s^{-1}$ (lower panel). Left, middle, and right columns display the first, second and third vertical modes. Black, red, and blue curves on each panel show $n=0, n=1, n=2$ modes, respectively. All modes represent the equatorially trapped waves.
\label{fig3}}
\end{figure*}

Eq. ~(\ref{Eq.(5)}) leads to the Bessel equation for the linear temperature profile (Eq. ~(\ref{Eq.6}))
\begin{equation}
x^2\frac{\partial ^2 V(x)}{\partial x^2} + x \frac{\partial V(x)}{\partial x}+ ( x^2 - n^2) V(x) = 0,
\label{Eq.(8)}
\end{equation}
where
   \begin{equation}
       x=2\sqrt{H_0-\epsilon z} \frac{\sqrt{\gamma (1-\epsilon )-1}}{\epsilon \sqrt{\gamma h}}
       \label{Eq.(9)}
       \end{equation}
and $n=(1-\epsilon)/\epsilon$.
Solutions of the equations are the Bessel functions of the order $n$, $J_n(x)$ and $Y_n(x)$. In Eq. ~(\ref{Eq.(8)}), there are two unknown parameters, $\epsilon$ and $h$, assuming that the surface temperature and hence $H_0$ is known. Then fixing the temperature gradient, $\epsilon$, leads to the determination of $h$ using boundary conditions. Both, $J_n(x)$ and $Y_n(x)$ are bounded for $z \rightarrow -\infty$ (i.e. towards stellar center). On the other hand, $Y_n(x) \rightarrow \infty$ for $x \rightarrow 0$ i.e. for $z \rightarrow H_0/ \epsilon$. Therefore, $J_n(x)$ seems to be a more appropriate solution of Eq. ~(\ref{Eq.(8)}). The free boundary condition (when the Lagrangian pressure is zero) at the surface $z=0$, that is at $x=x(z=0)$, leads to the equation \citep{Albekioni2023}

   \begin{equation}
      \frac{\partial J_n (x)}{\partial x} - \frac{1}{x}\frac{3(1-\epsilon)}{\epsilon} J_n(x) = 0.
      \label{Eq.(10)}
   \end{equation}
For each value of $\epsilon$, the free boundary condition, Eq. ~(\ref{Eq.(10)}), gives an infinite number of zeros and hence an infinite number of $h$. Each of the zeroes and hence $h$ corresponds to the particular vertical mode of waves.

Fig. (\ref{fig1}) shows the equivalent depth vs different temperature gradients for different zeroes of Eq. ~(\ref{Eq.(10)}). 

Only the first five zeros, hence the first five vertical harmonics, are shown in this figure in the temperature gradient range of 0.3-0.39 (note that $\epsilon=0.4$ is the upper limit of radiative medium). When the temperature gradient approaches the adiabatic limit ($\epsilon=0.4$), then the corresponding equivalent depth is reduced for all five vertical harmonics. We need to keep in mind that each calculated equivalent depth must lead to bounded solutions at poles when one inserts it in Eq. ~(\ref{Eq.(4)}).
This upper limit of the equivalent depth, $h_c$, is shown by black lines in Fig.(\ref{fig1}) for different rotation rates of stars (Note, that here solar angular frequency has a value $\Omega_{\odot}=3 \times 10^{-6} s^{-1}$). All the solutions with $h>h_c$ do not satisfy polar boundary conditions, therefore are invalid for our consideration (see the next section). It is seen that the rapidly rotating stars allow all vertical harmonics in considered intervals of the temperature gradient to be bounded at poles. On the other hand, for the slowly rotating stars (similar to the Sun) only fifth and higher overtones are under the limit for the whole range of temperature gradient.  

Next, we analyze the vertical structure of individual vertical modes and its dependence on the value of temperature gradient. We assume two different values of $\epsilon$: 0.35 and 0.39. The first three values of equivalent depth corresponding to the first three vertical modes for $\epsilon=0.35$ calculated from Eq. ~(\ref{Eq.(10)}) are $h=0.0485 \, H_0, 0.0193  \, H_0, 0.0104  \, H_0$, respectively. For $\epsilon=0.39$, the corresponding values are $h=0.0077 \, H_0, 0.0032 \, H_0, 0.0018 \, H_0$. The corresponding solutions of Eq.(\ref{Eq.(5)}), which is the latitudinal velocity of the Rossby waves, are plotted Fig. (\ref{fig2}). 
The figure shows that the wavelength is shorter for the higher modes as expected, but the higher modes penetrate slightly deeper in the interior. On the other hand, a larger temperature gradient yields a slightly longer wavelength of the waves as seen from the comparison of upper and lower rows.\footnote{Note, that the lower left panel of Figure 2 of this paper is the same as the right panel of Fig.1 in \citet{Albekioni2023}. However, the amplitudes at $z=0$ are different in these two plots, as here we used the same amplitudes for all six plots for better comparison. This different normalisation obviously does not affect the vertical structure of the waves.}

\begin{figure*}
\centering
\includegraphics[width=17cm]{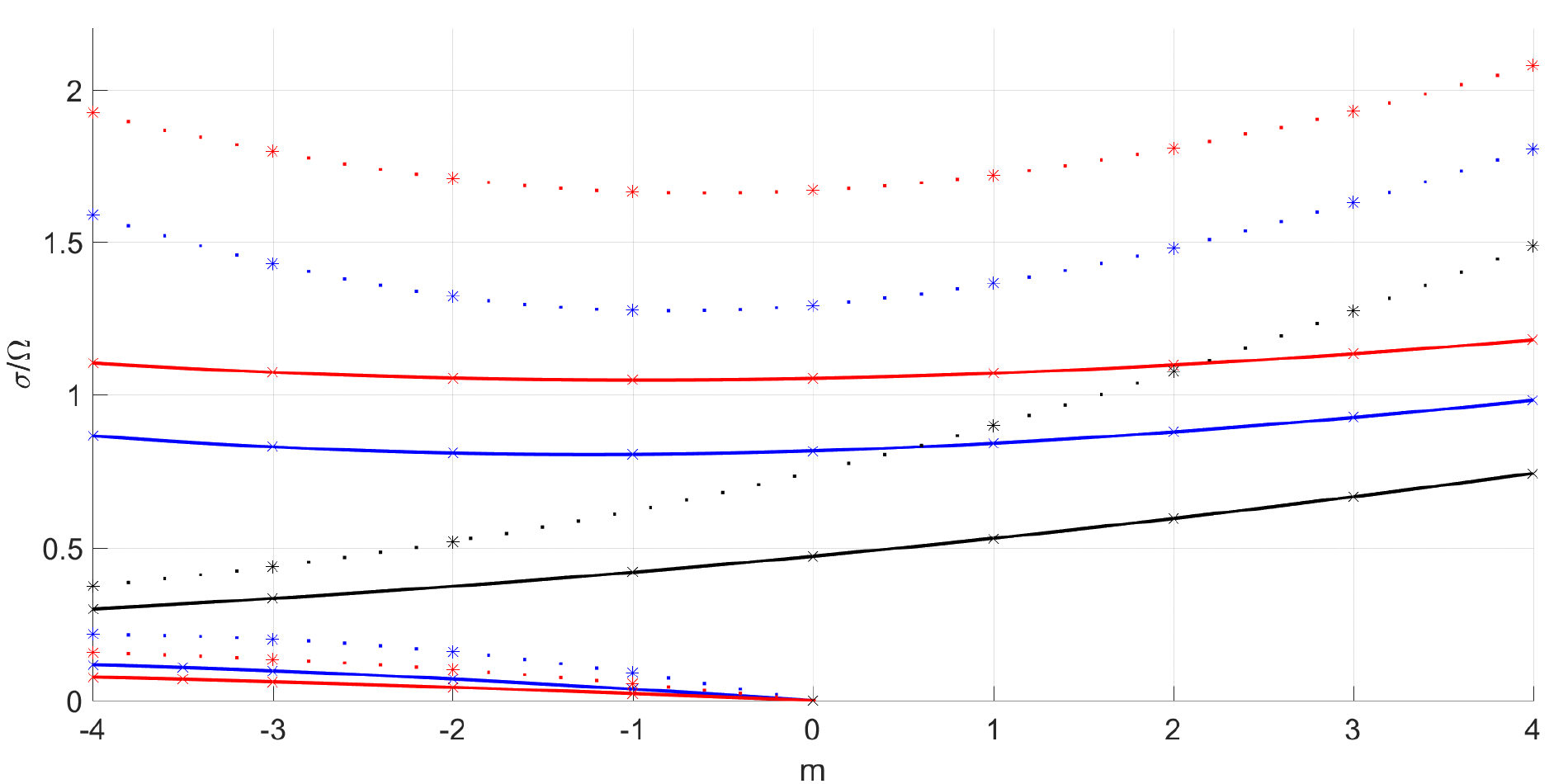}
\caption{Frequencies of Rossby (lower curves), Rossby-gravity (middle curves with black color) and inertia-gravity waves (upper curves) vs toroidal wavenumber, $m$, obtained from Eq. \ref{Eq.(14)}. Black, blue and red colors show the modes with $n=0
$, $n=1$, and $n=2$, respectively. The frequencies match with the first vertical modes corresponding to the temperature gradients of $\epsilon=0.39$ (solid lines) and $\epsilon=0.35$ (dashed lines) for the stellar angular frequency of $3\Omega_{\odot}$.
\label{fig4}}
\end{figure*}

\begin{figure*}
\centering
\includegraphics[width=17cm]{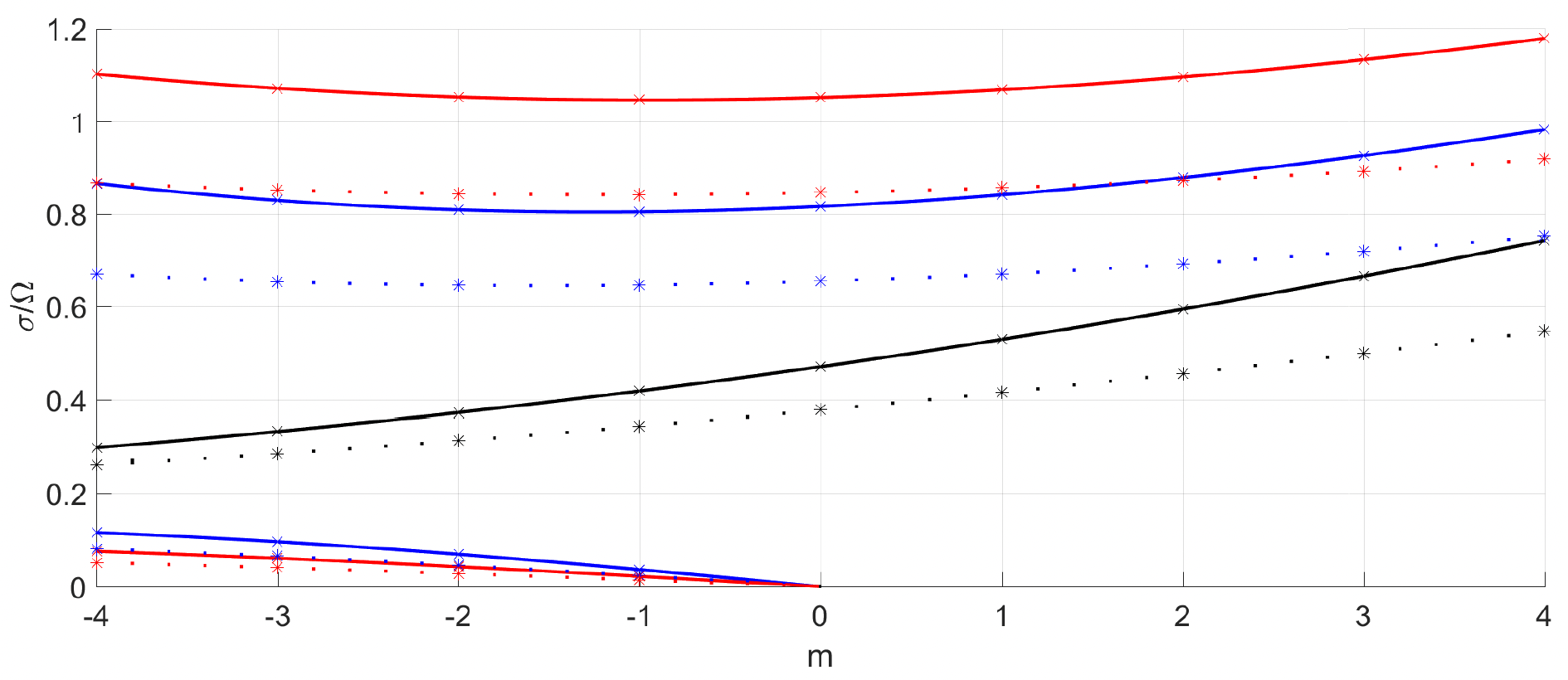}
\caption{Frequencies of Rossby (lower curves), Rossby-gravity (middle curves with black color) and inertia-gravity waves (upper curves) vs toroidal wavenumber, $m$, obtained from Eq. \ref{Eq.(13)}. Black, blue and red colors show the modes with $n=0
$, $n=1$, and $n=2$, respectively. The frequencies match with the first (solid lines) and the second (dashed lines) vertical modes for the temperature gradient of $\epsilon=0.39$ for the stellar angular frequency of $3\Omega_{\odot}$.
\label{fig5}}
\end{figure*}

\begin{figure*}
\centering

\includegraphics[width=17cm]{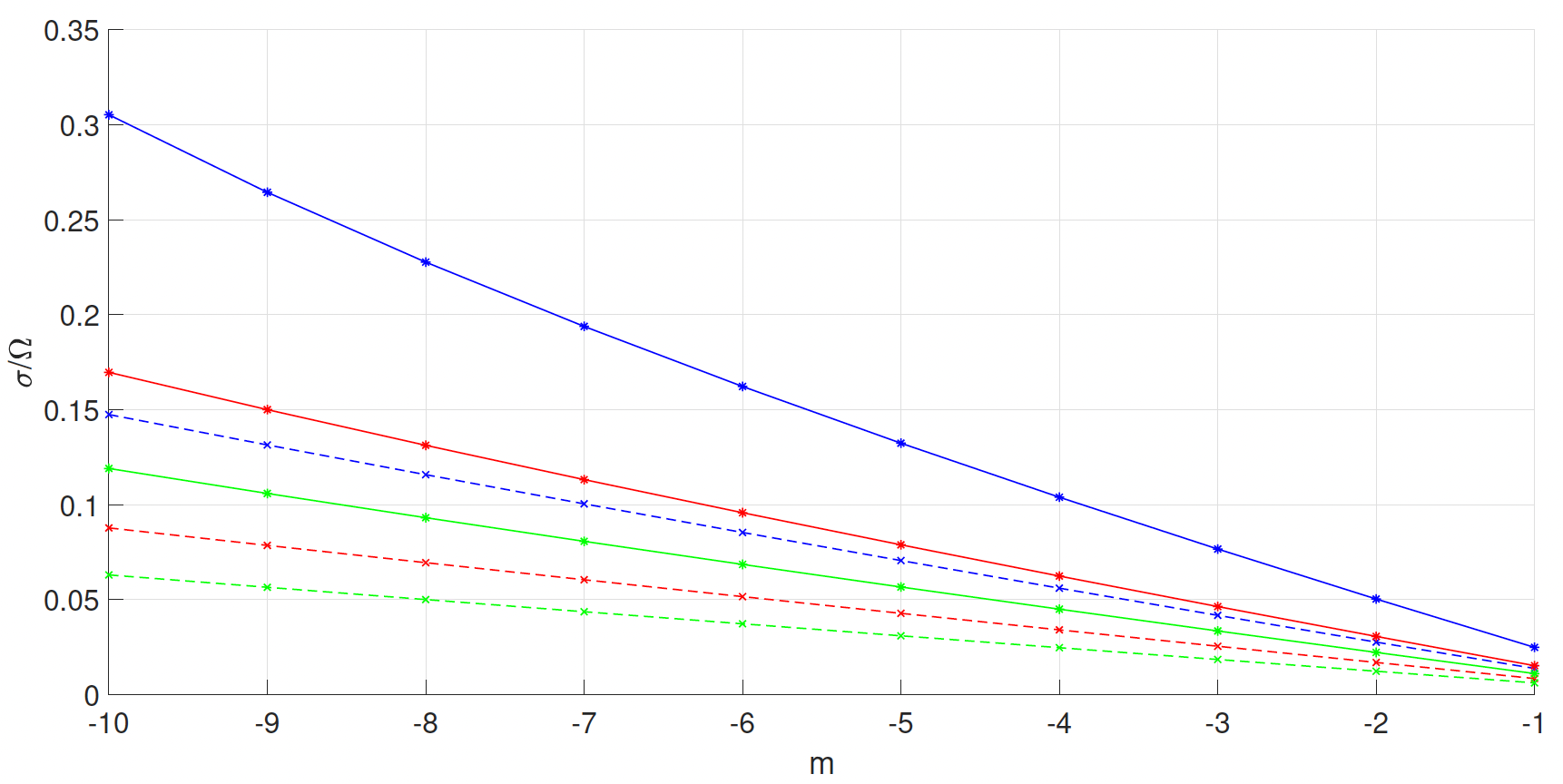}

\caption{First (blue), second (red) and third (green) vertical harmonics of Rossby waves vs toroidal wavenumber, $m$, obtained from Eq. \ref{Eq.(13)} for the temperature gradient of $\epsilon=0.39$ and the stellar angular frequency of $3\Omega_{\odot}$. Solid and dashed lines correspond to the modes with $n=2$ and $n=4$, respectively. 
\label{fig6}}
\end{figure*}

\section{Latitudinal structure of Rossby waves}\label{sec4}

The obtained equivalent depths, $h$, which correspond to the different vertical modes of Rossby waves, must be inserted in Eq.(\ref{Eq.(4)}) and the solutions satisfying the bounded conditions at poles must be found. Consequently, only the vertical modes that lead to the latitudinal solutions vanishing towards the poles are valid. The latitudinal structure of the waves described by Eq.(\ref{Eq.(4)}) crucially depends on the equivalent depth. It is found by \citet{Albekioni2023} that the sufficiently small equivalent depth results in the solutions, which are concentrated around the equator. The equivalent depth in Figure \ref{fig1} has a small value for almost all modes for rapid rotations ($\Omega>\Omega_{\odot}$), therefore, almost all vertical modes represent equatorially trapped waves. Near the equator, Coriolis parameter can be expanded as $f=\beta y$, where $\beta=2\Omega/R$ and the Eq.(\ref{Eq.(4)}) is transformed into parabolic cylinder equation
  \begin{equation}
        \frac{\partial^2 \Psi}{\partial y^2} + \left[-\frac{k \beta}{\sigma}   +\frac{\sigma^2 - k^2 c^2}{c^2} - \frac{\beta^2 y^2}{c^2} \right ] \Psi(y) = 0,
        \label{Eq.(11)}
        \end{equation}
where $c=\sqrt{g h}$ is the surface gravity speed. When the Lamb parameter of particular wave mode with the equivalent depth $h$, $\varepsilon=4\Omega^2 R^2/gh$, is much larger than unity, then the solutions of Eq.(\ref{Eq.(11)}) are trapped in low latitudes and exponentially vanish towards poles. This happens for small equivalent depth, $h\ll H_0$, and fast rotation, $\Omega > \Omega_{\odot}$. In this case, the solutions of Eq.(\ref{Eq.(11)}) are in terms of Hermite polynomials (see details in \citet{Albekioni2023}) 
\begin{equation}
\Psi=\Psi_0\exp{\left (-\frac{\sqrt{\varepsilon}}{2}\frac{y^2}{R^2}\right )}H_n\left (\varepsilon^{1/4}\frac{y}{R} \right ),
        \label{Eq.(12)}
        \end{equation}
and the waves are governed by the dispersion equation 

\begin{equation}
\omega^3 - \frac{4}{\sqrt{\varepsilon}} \left (2n+1 + \frac{m^2}{\sqrt{\varepsilon}}\right )\omega - \frac{8m}{{\varepsilon}} = 0,
        \label{Eq.(13)}
        \end{equation}
where $\omega=\sigma/\Omega$ is the normalised wave frequency, $m=kR$ is the normalised toroidal wavenumber (can be assumed as discrete $m=0,1,2...$), and $n=0,1,2...$ denotes the number of zeroes along the latitudinal direction. $n=0$ means that the mode has no zero along the latitudinal direction i.e. it corresponds to the sectoral harmonic, $n=1$ has one zero at the equator, etc. For $n\geq 1$, the dispersion relation governs two higher frequency prograde and retrograde inertia-gravity waves and one lower frequency retrograde Rossby wave. For $n= 0$, Eq. (\ref{Eq.(13)}) governs prograde and retrograde Rossby-gravity waves.

\begin{figure*}
\centering

\includegraphics[width=18cm]{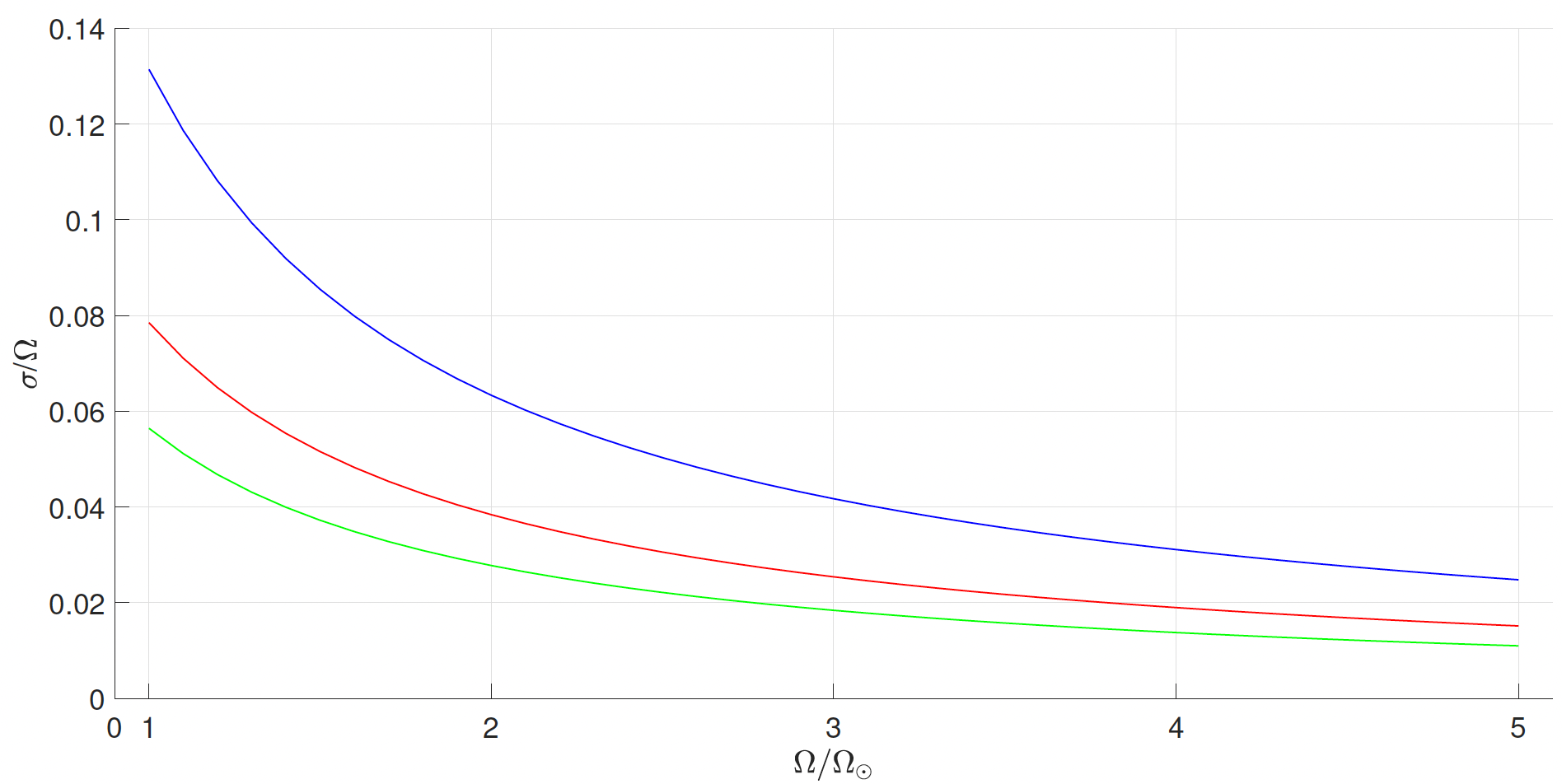}

\caption{Frequency of different vertical modes of Rossby wave vs stellar rotation for a vertical temperature gradient of $\epsilon=0.39$.  Blue, red, and green lines correspond to the first, second, and third vertical modes, respectively. Here the toroidal and latitudinal wavenumbers are $m=-1$ and $n=1$, respectively. \label{fig7}}
\end{figure*}

Eq. (\ref{Eq.(12)}) shows that the latitudinal structure of the waves is governed by $\varepsilon$: larger $\varepsilon$ leads to the
stronger decrease in the wave amplitude. One can estimate the critical $\varepsilon$ from Eq. (\ref{Eq.(12)}) as $(\sqrt{\varepsilon} y^2/2 R^2)_{y=\pi R/2}=1$, which yields $e$-times decrease of polar values with regards to the equator. This gives the critical value of $\varepsilon$ as $\approx 25$. All modes with $\varepsilon > 25$ vanish towards poles, therefore they are valid solutions, while the modes with $\varepsilon < 25$ are not bounded at poles. Consequently, for each angular frequency, $\Omega$, there is the critical value of the equivalent depth, $h$, which defines valid solutions of Rossby waves (here we assume fixed $R$ and $g$). The critical equivalent depths for different angular velocities of stars are plotted by horizontal black lines in Fig.(\ref{fig1}). 

The upper panels of Fig. (\ref{fig3}) show the latitudinal structure of the first three vertical modes in the case of the temperature gradient $\epsilon=0.39$ and the angular frequency of $\Omega=3\Omega_{\odot}$. It is seen that the waves are located between $\pm 50^0$, therefore they are equatorially trapped waves. Their amplitudes vanish towards the poles and hence satisfy the polar boundary conditions. The higher modes are more concentrated towards the equator. This happens because $\varepsilon$ is inversely proportional to the equivalent depth, hence the smaller $h$ yields the larger $\varepsilon$ in the case of fixed stellar angular frequency. Lower panels of Fig. (\ref{fig3}) display the structures of the same modes but for higher angular frequency $\Omega=5\Omega_{\odot}$. It is seen that the wave profiles are squeezed towards the equator in comparison to the slower rotation and the waves are trapped between $\pm 30^0$ latitudes. Therefore, the stars with faster rotation display stronger trapping of Rossby waves around the equator.  

\section{Frequency of the waves}\label{sec5}

Eq. (\ref{Eq.(13)}) shows that the wave frequency depends on the Lamb parameter, $\varepsilon$, and on the latitudinal wavenumber, $n$. On the other hand, $\varepsilon$ is proportional to the frequency of stellar rotation (and radius) and inversely proportional to the equivalent depth, $h$ (and the surface gravitational acceleration, $g$). As the equivalent depth is a function of the temperature gradient (see Figure 1) and has various values for different vertical modes, the wave frequency also depends on the two parameters. 

Fig. (\ref{fig4}) exhibits the solutions of Eq. (\ref{Eq.(14)}) with regards to $m$ for the different values of the temperature gradient.  Solid lines show the first vertical modes of Rossby, Rossby-gravity, and inertia-gravity waves for the temperature gradient of $\epsilon=0.39$. Lower (upper) blue and red curves display the first vertical modes of the Rossby (inertia-gravity) waves with a latitudinal wavenumber of $n=1$ and $n=2$, respectively. The middle black curve corresponds to the Rossby-gravity mode with $n=0$. The dashed curves show the same modes for the temperature gradient of $\epsilon=0.35$. The difference between the two temperature gradients is clear, so that the larger temperature gradient leads to the lower frequencies of all waves. 

Fig. (\ref{fig5}) presents the solutions of Eq. (\ref{Eq.(13)}) with regards to $m$ for the different vertical modes. Solid (dashed) lines show the first (the second) vertical modes of Rossby, Rossby-gravity, and inertia-gravity waves for the temperature gradient of $\epsilon=0.39$. Lower (upper) blue and red curves display the Rossby (inertia-gravity) waves with a latitudinal wavenumber of $n=1$ and $n=2$, respectively. The middle black curves correspond to the Rossby-gravity mode with $n=0$. The difference between the various vertical modes is distinct so that the higher vertical modes lead to lower frequencies of all waves. Fig. (\ref{fig6}) presents the solutions of Eq. (\ref{Eq.(13)}) with regards to $m$ for the different vertical modes of Rossby waves in the case of the temperature gradient $\epsilon=0.39$ for $3\Omega_{\odot}$. Here the latitudinal wavenumbers are $n=2$ and $n=4$ and displayed by solid and dashed lines, respectively. The frequency difference of different vertical modes is clear.

As $\varepsilon$ is also a function of angular frequency, then the wave frequency depends on the stellar rotation.  Fig. (\ref{fig7}) presents the frequency of various vertical modes of Rossby wave vs the stellar rotation. The Rossby wave frequency is inversely proportional to $\varepsilon$, which means that the wave frequency normalized by the stellar angular frequency is decreasing for faster-rotating stars, especially for the higher vertical modes. It is also seen that the difference between the frequencies of different vertical modes decreases for faster-rotating stars. 

\begin{figure*}
\centering
\includegraphics[width=18cm]{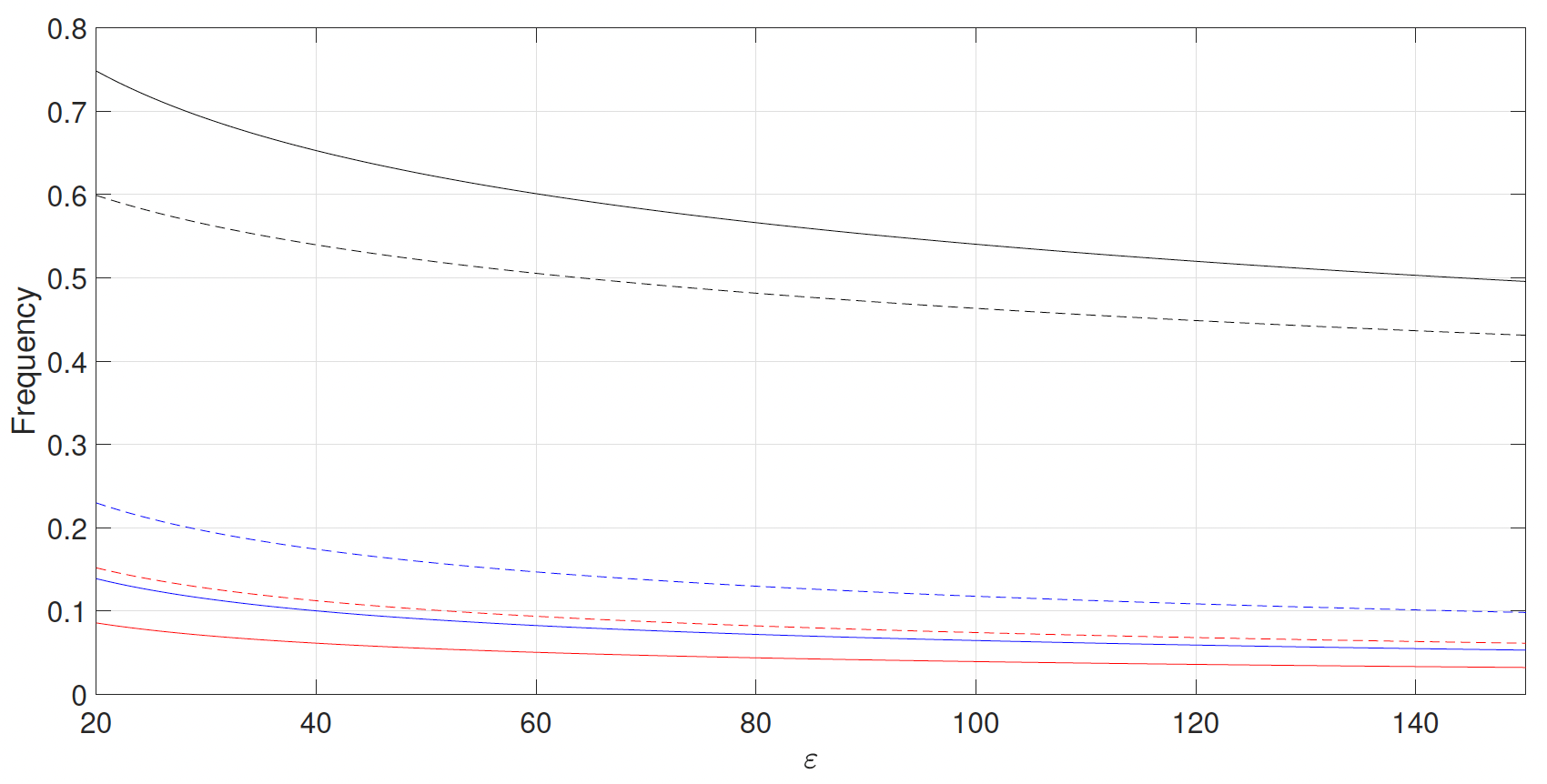}
\caption{Frequency of Rossby waves vs $\varepsilon=4\Omega^2 R^2/gh$. The black solid (dashed) curve corresponds to the Rossby-gravity mode with $(n,m)=(0,-1)$ ($(n,m)=(0,-2)$). The blue solid (dashed) curve corresponds to Rossby waves with $(n,m)=(1,-1)$ ($(n,m)=(1,-2)$). The red solid (dashed) curve corresponds to Rossby waves with $(n,m)=(2,-1)$ ($(n,m)=(2,-2)$). The vertical mode of the waves can be determined by corresponding equivalent depth, $h$, from Fig. (\ref{fig1}) fixing the vertical temperature gradient. \label{fig8}}
\end{figure*}

Fig. (\ref{fig8}) presents the solutions of Eq. (\ref{Eq.(13)}) vs $\varepsilon=4\Omega^2 R^2/gh$ for Rossby waves with with different toroidal ($m$) and latitudinal ($n$) wavenumbers. Equivalent depth for particular vertical modes and temperature gradient in the stellar near surface interior can be determined from Fig. (\ref{fig1}). Therefore, the observed frequency in light curves of a star with known rotation, surface gravity, and radius may determine the corresponding equivalent depth $h$ from Fig. (\ref{fig8}). Then Fig. (\ref{fig1}) can be used to find suitable vertical temperature gradient and vertical mode of Rossby waves as well as the appropriate vertical structure of the waves. This could be an important tool for future seismology of stellar interior by Rossby-type waves.

\section{Discussion and conclusion}

Growing data of stellar light curves obtained by recent space missions requires improving indirect tools for the estimation of internal parameters. Acoustic and gravity waves are obvious choices for asteroseismology, but it becomes increasingly clear that the Rossby waves also have significant potential to be used for probing stellar interiors \citep{Aerts2021}. The theory of stellar Rossby waves (or r-modes) was developed using expansions with a small parameter for high-order modes of rapidly rotating stars \citep{Papaloizou1978} and for low-order modes in slowly rotating stars \citep{Provost1981, Saio1982}. On the other hand, \citet{Lee1997} and \citet{Townsend2003} used traditional approximation to separate the horizontal and vertical equations in order to obtain the Laplace tidal equation for horizontal variations. Then they solved the Laplace tidal equations and obtained the dependence of wave frequency and separation constant. Then one can find the separation constant for each wave frequency and use it to solve the vertical equations to obtain the vertical structure of the modes. In the recent paper \citep{Albekioni2023}, we used the same formalism of separation, but first solved the vertical equations for the linear temperature profile in free surface boundary conditions and found the corresponding value of separation constant, which is actually an equivalent depth of Rossby waves. Consequently, we derived the oscillation spectrum of Rossby waves from the horizontal equations for the first vertical mode and fixed vertical temperature gradient. It was found that the equivalent depth derived from surface boundary conditions has a small value, therefore corresponding horizontal solutions are concentrated around the equator leading to the equatorially trapped waves. This solution justifies the use of Cartesian coordinates as the solutions and frequencies of equatorially trapped Rossby waves are identical in rectangular and spherical considerations \citep{Matsuno1966, Longuet-Higgins1968, Zaqarashvili2021}.  On the other hand, the difference in periods of various vertical modes, e.g. period spacing pattern, is often used to identify observed wave modes \citep{Aerts2021} and hence is very useful in asteroseismology. Consequently, it is of vital importance to study various vertical modes in the discussed formalism. Here we continue the similar study for various values of temperature gradient and vertical overtones of waves. 

\begin{figure*}
\centering
\includegraphics[width=18cm]{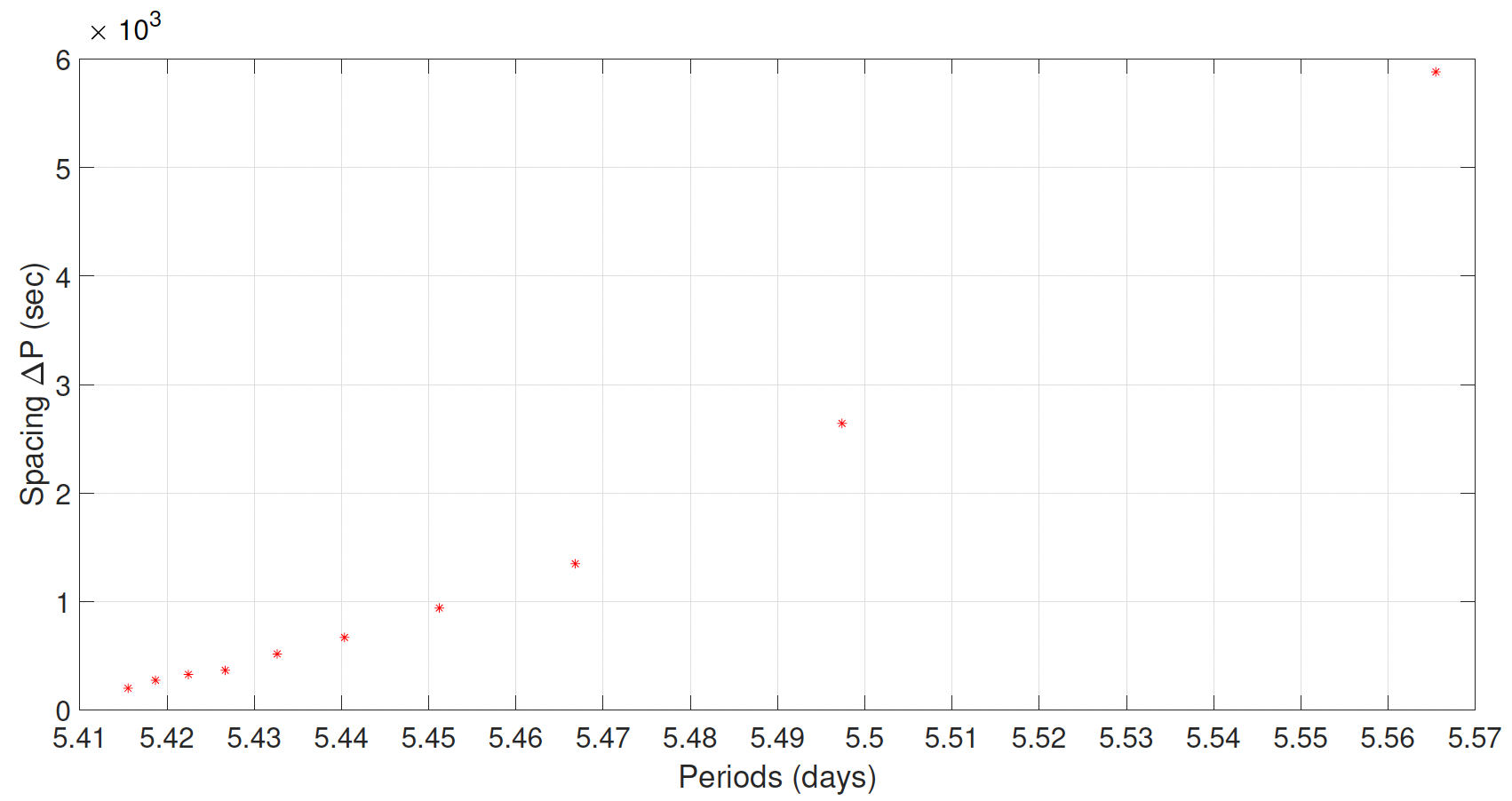}
\includegraphics[width=18cm]{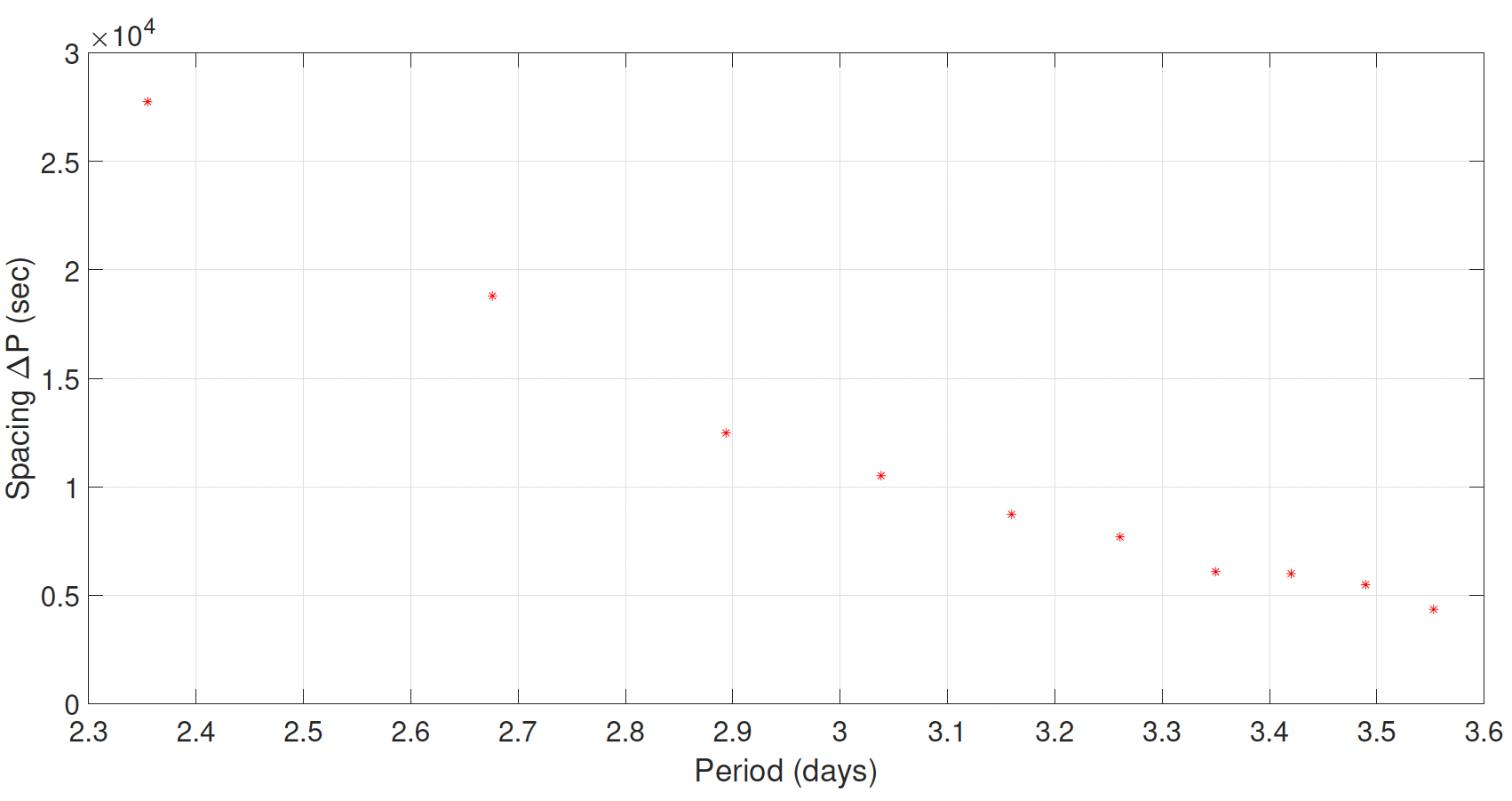}
\caption{Spacing ($\Delta P$) vs period patterns for Rossby (upper panel, $m=-1$ and $n=2$ mode) and inertia-gravity (lower panel, $m=1$ and $n=2$ mode) waves for the vertical temperature gradient of $\epsilon=0.39$ and the stellar angular frequency of $\Omega=5\Omega_{\odot}$ (i. e. the rotation period of $\sim$ 5.4 days). Wave periods are calculated in the inertial frame. The first ten vertical modes are used to construct the period-spacing patterns; the vertical order of the modes is increasing from left to right.  \label{fig9}}
\end{figure*}

Figure 1 shows the dependence of equivalent depth, which was obtained from the solution of vertical equations using free boundary condition, on vertical temperature gradient ($|dT/dz| \sim \epsilon$) for the first five vertical modes. The value of the equivalent depth decreases for higher vertical overtones and for larger temperature gradients. We note that $\epsilon=0.4$ corresponds to the adiabatic temperature gradient, therefore the radiative (or convectively stable) medium yields $\epsilon<0.4$.
The corresponding vertical structure of the modes is plotted in Figure 2 for different temperature gradients. We see that the vertical wavelength decreases significantly for higher overtones and imperceptibly for stronger temperature gradients. In all cases, the amplitude of latitudinal velocity decreases with depth so that the Rossby modes are concentrated near the surface layer with a thickness of 50 surface scale height. Higher overtones penetrate slightly deeper. It must be noted that the rotation kernels and vertical structure of $m=1$ Rossby mode in the recent paper by  \citet{VanReeth2018} showed only a slight decrease of the wave amplitude with depth in upper layers and subsequent growth toward the center of stars for solid body rotation. The difference between the results depicted in Figure 2 of \citet{VanReeth2018} and Figure \ref{fig2} in the current paper arises due to the density factor, $\rho_0$. In Figure 2 of our paper, the displayed velocity does not include density as indicated by the figure labels. But in Figure 2 of  \citet{VanReeth2018}, the rotation kernels incorporate density \citep{Aerts2010}, which leads to the mentioned discrepancy. If one plots the velocity with the factor of $\sqrt{\rho_0}$ in Figure 2 in our paper, then one can find only a gradual decrease in amplitude with depth in coincidence with \citet{VanReeth2018}. The absence of amplitude growth in deeper layers in Figure 2 is probably caused due to the used linear temperature gradient. The inclusion of differential rotation obviously will lead to new vertical profiles of the waves.

Small values of equivalent depth lead to the parabolic cylinder equation for horizontal direction with bounded solutions in terms of Hermite polynomials and with corresponding dispersion relation of Rossby, Rossby-gravity, and inertia-gravity waves. The latitudinal structure of the waves depends on the parameter $\varepsilon=4\Omega^2 R^2/g h$ as shown by Eq. (13). Larger  $\varepsilon$ leads to the stronger equatorial confinement of the waves. Figure 3 shows the latitudinal structure of the first three vertical modes for different rotation rates. The waves are strongly trapped near the equator for rapidly rotating stars: the waves are confined between latitudes of $\pm 30^0$ for 5 $\Omega_{\odot}$. In this paper, we consider the stars with outer radiative zones, which occupy the upper part of the Herzsprung-Russel diagram. These stars generally have rapid rotation compared to $\Omega_{\odot}$ due to the weak magnetic breaking \citep{Kraft1967}. Therefore, the stars considered in this paper are fast rotators, hence the waves are trapped around the equator for almost all modes and the value of the temperature gradient. 

The wave frequency strongly depends on $\varepsilon$, hence on the surface rotation frequency and corresponding equivalent depth. 
As the equivalent depth is a function of the temperature gradient, the wave frequency also hinges on it. Figure 4 shows that the frequency of all wave modes significantly depend on the gradient, which implies that observations may provide the estimation of the parameter. Wave frequency also significantly depends on stellar rotation as seen from Figure 7 for the Rossby node with $m=-1$ and $n=1$. The dependence seems more important at the slow rotation part of the interval but becomes imperceptible for rapid rotations. 
 
Observations of light curve variations may provide asteroseismic sounding of some parameters. Here we consider solid body rotation without magnetic field, therefore we can estimate the vertical temperature gradient. Incorporation of differential rotation \citep{Gizon2020} and the magnetic field \citep{Zaqarashvili2018} in the model may widen the application further to magnetic field strength and the differential rotation rate. The frequency of Rossby waves is smaller than the angular frequency of rotation, while the frequency of inertia-gravity waves is comparable to the rotation frequency (see Figures 4-5). It should be noted that the wave frequency is calculated in the rotating frame in this paper, therefore one needs to transform it in the inertial frame in order to compare with observations. Transformation of frequency occurs with $\sigma_{obs}=m \Omega+\sigma$. Hence, the Rossby waves with $m=-1$ will have a slightly lower frequency compared to the rotational frequency $|\sigma|<\Omega$, while the waves with $m=-2$ will have the frequency in the interval of $\Omega<|\sigma|<2\Omega$. 

We mentioned above that the vertical modes are important to construct the period-spacing patterns, which are used to compare the theory with observations. As an example, we plot the period-spacing patterns for Rossby and inertia-gravity waves on Fig. \ref{fig9}  for a specific value of the temperature gradient, $\epsilon=0.39$. We calculated the first ten vertical modes for the stellar angular frequency of $\Omega=5\Omega_{\odot}$ and the horizontal wavenumbers of $n=2$ and $|m|=1$. We see that the Rossby and inertia-gravity waves show expected behavior as shown in the literature \citep{Aerts2021}.
This analysis may serve as another valuable tool to facilitate comparisons between observations and theoretical predictions.

Observations of frequency in the light curves of stars can be compared with the theoretical spectra of the waves. Figure 8 displays the dependence of wave frequency on  $\varepsilon$ for several modes of Rossby waves with different $m$ and $n$. The observed frequency will then determine the corresponding equivalent depth $h$ if the radius, surface gravity, and rotational frequency of the star are known. Once the equivalent depth is estimated, one can use Fig.\ref{fig1} to identify the corresponding $\epsilon$ and vertical mode number of the waves. Consequently, one can estimate the vertical temperature gradient and the vertical structure of different harmonics. This methodology could potentially serve as an important tool for future asteroseismic studies of stellar interiors.

\textit{Acknowledgements}: MA was supported by Volkswagen Foundation and by Shota Rustaveli National Science Foundation Georgia (SRNSFG) [Grant number N04/46-9] in the framework of the project "Structured Education in Quality Assurance Freedom to Think". PB was supported by the Austrian Fonds zur F{\"o}rderung der Wissenschaftlichen Forschung (FWF) project P32958. This paper resulted from discussions at workshops of ISSI (International Space Science Institute) team (ID 389) "Rossby waves in astrophysics" organized in Bern (Switzerland).

\bibliography{Wiley-ASNA}%

\end{document}